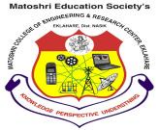
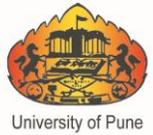

# *Recommendation System for Outfit Selection (RSOS)*

Shiv H. Sutar and Akshata H. Khade

**Abstract:** We propose a system which will be able to recommend the user to choose appropriate outfits suits to their personality. The necessity of this system is to reduce the outfit selection and purchasing time; this will also help to create tailor made outfits as per the personality traits.  The guidelines for selection of their respective outfits are based upon various bodily parameters that evolve with the learning of available labeled and unlabeled data. The system is based on two modules of processes; first one is to recognize the features for usage of outfits like traditional, western, functional, daytime or night etc, second is to calculate the body measurement parameters. The proposed system will have image capturing by using HAAR feature or input device for getting body parameters. We intend to classify and extract the best possible outfits from the system by using HIGEN MINER algorithm. The applications of outfit selection will be ranging from manual gender selection, image processing with body feature extractions, Value comparison with database by using different statistical techniques and data mining algorithms. After that it will recommend best outfits as per body parameters, inputs and availability

## I. INTRODUCTION

Online shopping is a form of electronic commerce which allows consumers to directly buy services from a seller over the internet by using a web browser. The recent survey conducted by AC Nielson on global online consumers say that clothing will continue to top the list for planned online purchases in the next six months.

Given the lack of ability to inspect merchandise before purchase, customers are at higher risk of fraud than face-to-face transactions also the cost of hidden charges, transit damage and higher cost of replacement.

The system is based on modules of two processes; first one is to recognize the body parameters by using image capturing device from fixed distance and also to get an input from the user. And second is to process image to get measurement or to calculate the body measurement parameters which will be used to mine the data from the database by using association rule. The system will show or recommend the best possible outfits based on the input parameter combinations. We will be studying different parameters buying behaviors and process to improve the project scope.

The survey done by Nielsen reports that 46 percent of global consumers said they purchased books in the past three months and 41 percent bought clothing online [12].

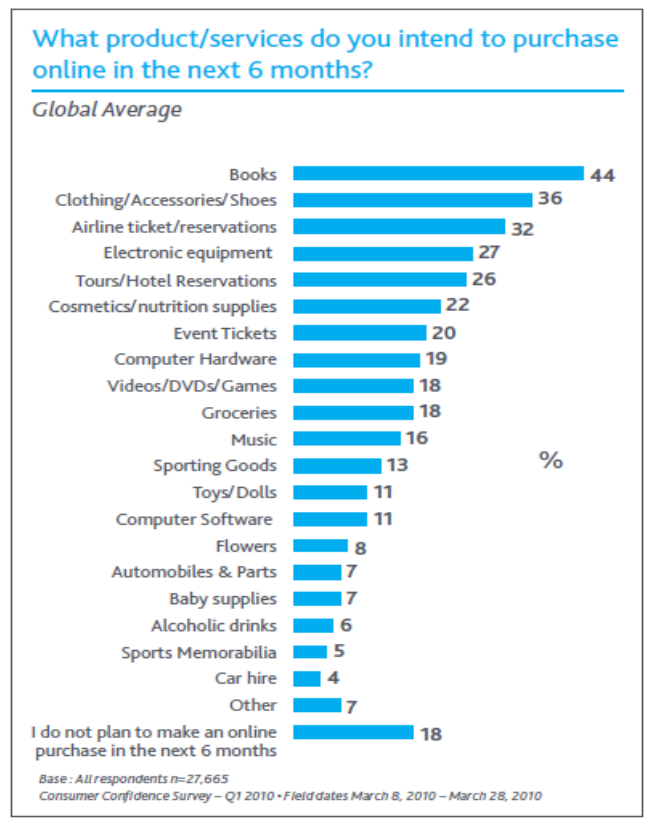

Fig. 1 Global Trend on Online Shopping [12].

## A. Recommendation System:

Recommendation System are a subclass of information filtering system that seek to predict the 'rating' or 'preference' that user would give to an item, they had not yet considered. Recommendation systems typically produce a list of recommendations in one of two ways - through collaborative or content-based filtering.

Collaborative filtering approaches is used to build a model from a user's past behavior (items previously purchased or selected and/or numerical ratings given to those items) as well as similar decisions made by other users, then use that model to predict items (or ratings for items) that the user may have an interesting.

Content-based filtering approaches utilize a series of discrete characteristics of an item in order to recommend additional items with related properties. These approaches are often combined.

A collection of outfits to motivate users personal taste and make the most of your wardrobe, this means you select the outfits based on your culture, personality, profession, environment etc.

It means to build or developed a system which will be able to help or guide the user to choose appropriate outfits suits to their personality. The necessity of this system is to reduce the outfit selection and purchasing time; this will also help to recommend outfits as per the personality traits. The guidelines for selection of their respective outfits are based upon various bodily parameters that evolve with the learning of available labeled and unlabeled data.

This paper is organized as follows. Section II discusses a review of earlier works. Section III Implementation details. In Section IV, we discuss result analysis and Section V concludes this paper.

## II. LITERATURE SURVEY

The proposed system is a robust real-time embedded platform to monitor the loss of attention of the driver during day and night driving conditions. Here the face is detected using Haar-like features and is tracked using a Kalman filter. The percentage of eye closure has been used to indicate the alertness level. The drawback of this paper is detection accuracy is low when lighting conditions are extremely dark or bright. [1].

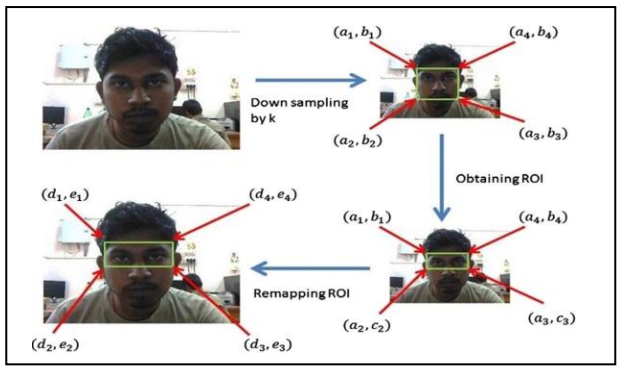

Fig. 2 Scheme of Eye Detection using Downsampling, Extraction, and Remapping of ROI [1].

Here proposes a dynamic pattern, namely the HIGEN i.e. HIstory GENeralized Pattern that represents the evolution, by reporting the information about its frequent generalizations characterized by minimal redundancy of an itemset in consecutive time periods, in case it becomes infrequent in a certain time. This performed on both real and synthetic datasets show the efficiency and the effectiveness of the proposed approach and its usefulness in a real application. The drawback of this paper is The HIGEN MINER algorithm automatically selects the generalized itemsets subsets at the cost of a higher number of dataset scans [2].

This paper proposes two algorithm namely utility pattern growth i.e. UP-Growth and UP-Growth+, for mining high utility itemsets with a set of effective strategies for pruning candidate itemsets. The high utility itemsets is maintained in a tree-based data structure i.e. utility pattern tree (UP-Tree) such that candidate itemsets can be generated efficiently with only two scans of database. The performance of both is compared with the state-of-the-art algorithms. The drawback of this paper is an additional database scan is performed after potential high utility itemsets are found for identifying their utilities. [3].

This paper proposes a system which will be choose appropriate sized outfits based upon users body parameters. Using this system to reduce personal presence in today's busy world for selection and measurements of outfits. The system is based on two processes:
1. To recognize the body parameters.
2. To get measurements or to calculate the body parameters like tailor measurements.

The drawback of this paper is to achieve more specific result, need to consider more parameters, but if increased no of parameters record set will be much longer [8].

This system contains a Virtual Assistant which is capable of finding a specific piece of garment for a specific occasion and in compliance with physical characteristics of the users. Here also described some scenario based design idea that is GUI. The representation of garment is described below fig 3 the attributes considered are fabric, style color, fit, pattern, season, gender, age, category etc [17].

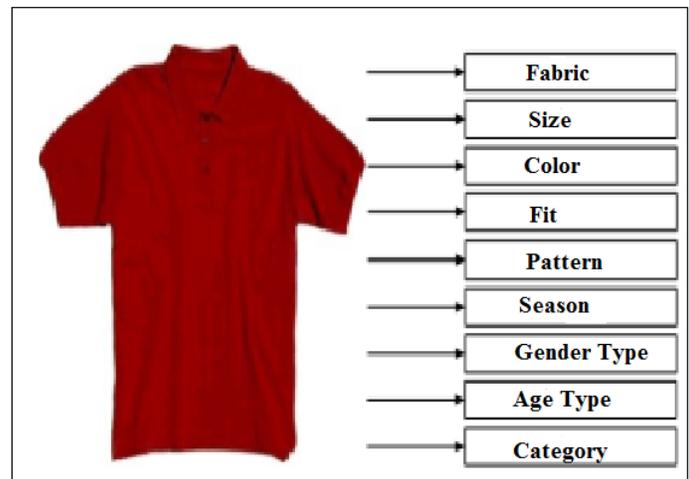

Fig. 3 Attributes to Describe a Garment Model [17].

Mining of frequent itemset is a widely exploratory technique that focuses on discovering recurrent correlations. The discovery of frequent generalized itemsets:
1. Provide a high-level abstraction of the mined Knowledge,
2. Frequently occur in the source data.

This system is expensive in its current states.

### III. IMPLEMENTATION DETAILS

A. *Block Diagram of RSOS:*

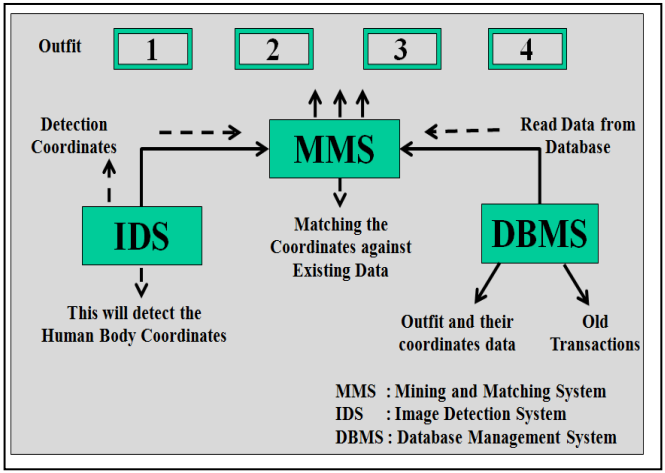

Fig. 4 Block Diagram of RSOS

The proposed system will have three basic modules Image Detection System, Mining and Matching System and Database Management System.

*Image Detection System:*

The IDS will have an image capturing device which will capture the input image for tracing out the body parameters by using HAAR features, also to have manual input like financial budget and profession and so on.

The HAAR feature can be defined as Sum of all pixel values within a rectangular area in an image. The sum of pixel values refers to the value in some two–dimensional representation of an image. There are some benefits of HAAR feature are this is very fast, easy to implement, feature value changes continuously when moving feature. Most attractive property is to speed of feature value evaluation. Here sliding window technique is used. And all Operation is done in constant time when integral image is created.

*Integral Image:*

Integral image means the simple rectangular features of an image are calculated using an Intermediate representation of an image. Main purpose of this is to speed up the process of evaluation of HAAR feature [15].

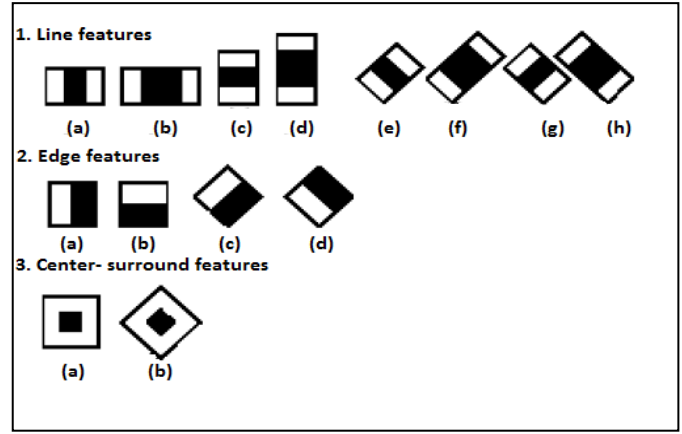

Fig. 5 Common HAAR features [15]

Fig. 5 2(e) requires another intermediate representation called the "Rotated Integral Image" or "Rotated sum Auxiliary Image".

*Training classifiers for facial features:*

Haar classifier cascades first be trained for Detecting human facial features, like mouth, nose, and eyes. For training of the classifiers two set of images are required. One set contains an scene or image that does not contain any object, This set of images is negative images. And in positive images, contain one or more instances of the object. The location of the objects within the positive images is specified by: image name, height, width and the upper left pixel [15].

Three separate classifiers were trained, for the eyes, for the nose, and for the mouth. Once, they were used to detect the facial features within another set of images from the database after the classifiers were trained [15].

Table 1 Accuracy of classifiers [15]

| Facial Features | Positive Hit Rate | Negative Hit Rate |
|---|---|---|
| Eyes | 93% | 23% |
| Nose | 100% | 29% |
| Mouth | 67% | 28% |

From literature survey, we found that HAAR feature is used in the system which is developed for Vision-Based System for Monitoring the Loss of Attention in Automotive Drivers. Basically HAAR feature is used for face detection, but in our case it will be used for body parameter recognition. HAAR object detection are color independent, fast, accurate and low false positive rate.

*Body Parameter Recognition :*

Colors (hair, eye, skin etc.) and shapes (bust, waist, neck, shoulder etc.) are used for body features recognition. Category Color and shape are used for determining a personal color palette as well as for clothes to fit [17].

Following parameters like Bust, Waist, Hips, Back-Width, Front Chest, Shoulder, Sleeve, Wrist, Upper Arm, Calf, Ankle, Nape- Waist, Front shoulder- Waist, Outside Leg will be extracted from the given image as a input. Body parameters are discovered by using HAAR feature.

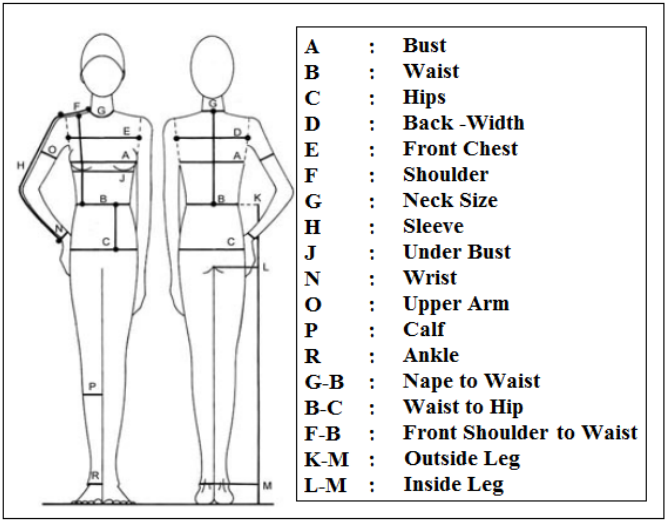

Fig. 6 Measurement points of the Body [17]

*2. Mining and Matching System:*

In this module, the captured image parameters by using HIGEN MINER algorithm will be used along with different manual inputs with different possible combinations to show the user the available outfits. The outfits will be mined from the linked database. Fig. 7. shows a simple hierarchy defined on the budget attribute. Table 4 reports the set of all possible frequent not generalized and generalized itemsets, and their corresponding absolute support values i.e. observed frequencies, mined from the datasets D1 and D2, and minimum support threshold is equal to 2. The readers can notice that some of the itemsets satisfy the support threshold in both D1 and D2, while the others are frequent in only one out of two months [2].

Table 2 Running Example Datasets

| Date | Profession | Budget | Outfit |
|---|---|---|---|
| 2012-10-01 | Engineer | 2500 | T-shirt |
| 2012-10-06 | Businessman | 2800 | T-shirt |
| 2012-10-08 | Teacher | 5800 | Jacket |
| 2012-10-20 | Teacher | 5800 | Jacket |
| 2012-10-23 | Doctor | 5200 | Jacket |

(a) DatasetD1.Product sales in October 2012

| Date | Profession | Budget | Outfit |
|---|---|---|---|
| 2012-11-03 | Engineer | 2500 | T-shirt |
| 2012-11-09 | Engineer | 2500 | T-shirt |
| 2012-11-15 | Teacher | 5800 | Jacket |
| 2012-11-28 | Doctor | 5200 | Jacket |
| 2012-11-29 | Businessman | 2800 | Jacket |

(b) DatasetD2.Product sales in November 2012

Consider, for instance, the itemsets {T-shirt, 2500} and {Jacket, 5800}. The former itemset is infrequent in D1 i.e. its support value in D1 is lower than the support threshold and becomes frequent in D2 due to a support value increase from October to November, while the latter shows an opposite trend. The fact that the generalized itemset {T-shirt, Medium} (frequent in both October and November 2012) has a specialization (i.e., {T-shirt, 2500}) that is infrequent in October while becomes frequent in the next month may be deemed relevant for decision making and thus, might be reported as it highlights a temporal correlation among spatial recurrences regarding product T-shirt. likewise, the information that, although {Jacket, 5800} becomes infrequent with respect to the minimum support threshold moving from October to November, its generalization {Jacket, High} remains frequent in both months could be relevant for analyst decision making as well [2].

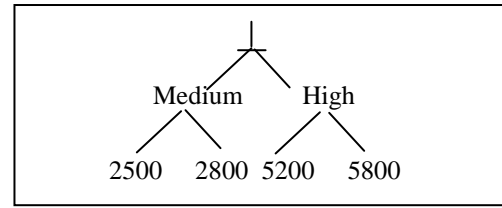

Fig. 7 Generalization Hierarchy for the Budget
Attribute in Example Datasets D1 and D2

Exploiting taxonomy over data items, items are aggregated into higher level i.e. generalized ones. Tables 2a and 2b report two example datasets, associated, with two consecutive months i.e. October 2012 and November 2012 [2].

Table 3 Extracted patterns

| Not Generalized Itemset | Sup D1 | Sup D2 |
|---|---|---|
| {2500} | 1(Inf.) | 2 |
| {5800} | 2 | 1(Inf.) |
| {T-shirt} | 2 | 2 |
| {Jacket} | 3 | 3 |
| {T-shirt, 2500} | 1(Inf.) | 2 |
| {Jacket, 5800} | 2 | 1(Inf.) |
| **Generalized Itemset** | **Sup D1** | **Sup D2** |
| {Medium} | 2 | 3 |
| {High} | 3 | 2 |
| {T-shirt, Medium} | 2 | 2 |
| {Jacket, High} | 3 | 2 |

a. Generalized and Not Generalized itemsets Mined from D1 and D2 min_sup =2

| HIGEN from D1 to D2 |
|---|
| {T-shirt}[sup=2] ~> {T-shirt}[sup=2] |
| {Jacket}[sup=3] ~>{Jacket}[sup=3] |
| {5800} [sup=2] ↗ {High} [sup=2] |
| {T-shirt, Medium}[sup=2] ↘ {T-shirt,2500}[sup=2] |
| {Jacket, 5800}[sup=2] ↗ {Jacket, High}[sup=2] |
| {Medium}[sup=3] ↘ {2500}[sup=2] |

b. Extracted HIGENs

In Table 3 the set of HIGENs, mined from the example dataset by enforcing an absolute minimum support threshold is equal to 2, is reported. The reference itemset is written in boldface. The relationship ↗ means that the right-hand side itemset is a generalization of the left-hand side one, ↘ implies a specialization relationship, while ~> implies that no abstraction level change occurs. For instance, {Jacket, 5800} ↗ {Jacket, High} is a HIGEN stating that, from October (D1) to November (D2) 2012, the not generalized itemset {Jacket, 5800} becomes infrequent with respect to the minimum support threshold while its upper level generalization (see in Fig. 7) remains frequent. All the HIGENs reported in Table 3b refer to a different not generalized reference itemset [2].

*3. Database Management System:*

DBMS will have system which will be used to pre-store the data on outfits based on the body parameter. The database will be used for data mining based on the input parameters.

IV. RESULT ANALYSIS

*A. Data Set:*

Data set for Female Jeans/ Legging:

Table 4 Data Set for Female jeans or Legging

| Sr. No | Name | Waist | Hips | Calf | Ankle | Outside Leg |
|---|---|---|---|---|---|---|
| 1 | L | 27 | 35 | 13 | 10 | 42 |
| 2 | M | 25 | 35 | 13 | 12 | 41 |
| 3 | N | 28 | 34 | 15 | 15 | 43 |
| 4 | O | 33 | 41 | 21 | 15 | 44 |
| 5 | P | 33 | 43 | 23 | 15 | 39 |

Data set Female top/ kurta/ one piece:

Table 5 Data set Female top/ kurta/ one piece

| Sr. No | Name | Bust | Waist | Hips | Back Width | Front Chest | Shoulder | Sleeve | Wrist | Nape to Waist | Front Shoulder to Waist |
|---|---|---|---|---|---|---|---|---|---|---|---|
| 1 | L | 30 | 29 | 37 | 19 | 12 | 5 | 4 | 9 | 16 | 15 |
| 2 | M | 31 | 30 | 41 | 20 | 13 | 6 | 5 | 10 | 17 | 16 |
| 3 | N | 32 | 31 | 42 | 17 | 14 | 4.4 | 2.6 | 10 | 15.6 | 14.6 |
| 4 | O | 33 | 33 | 33 | 18 | 12 | 6 | 0 | 14 | 17 | 15.6 |
| 5 | P | 29 | 27.5 | 46 | 18 | 15 | 4.9 | 10 | 15 | 15.8 | 14.6 |

*B. Result Set:*

Detect whole body parameters by using HAAR feature like face detection.

The RSOS system will suggest best outfit as per body parameters and also basis of availability. E.g. user wants western outfits

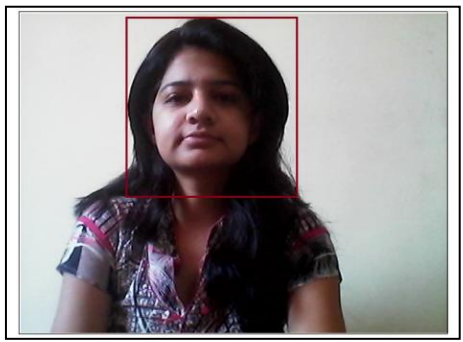

Fig. 8 Detection of Face

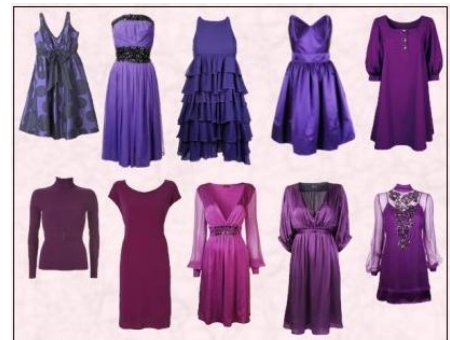

Fig. 9 Recommend Outfits

## V. CONCLUSION AND FUTURE WORK

We propose a system which will be able to recommended or guide the user to choose appropriate clothes or outfits suits to their personality. The guidelines for selection of their respective outfits are based upon various bodily parameters.

It is important to get feedback and opinions of different Retail outlets, Cloth designers, Customers and Sales Executive. Once we understand the buying behaviors and processes of consumer it will help us to broaden the scope of project.

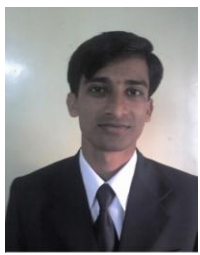
Shiv H. Sutar, received the Master's degree from University of Pune city Pune in Information Technology and Bachelor's degree in Computer Science from Swami Ramanand Tirth Marathawada University, city Nanded, state Maharashtra, country India. Currently he is working as a Assistant Professor in MIT college of Engineering, Pune. With 08 years of experience in Academics & Research. His area of interest is Systems Programming, Wireless Sensor Network, Compilers

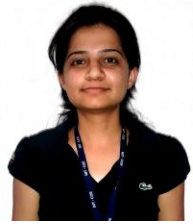
Akshata H. Khade, received the Bachelor's degree from University of Pune in Computer Engineering from VPCOE, Baramati, state Maharashtra, country India. She is currently pursuing full-time ME computers from MITCOE, university of pune. She has published one International Journal Papers and her research areas include Image Processing and Data Mining.